\documentclass[traditabstract]{aa}

\usepackage{epsfig,graphicx,natbib}

\begin{document}

\title{Investigating the nature of the Fried Egg nebula:}
\subtitle{CO mm-line and optical spectroscopy of IRAS~17163--3907}
\author{Sofia H. J. Wallstr\"om \inst{1}
\and S. Muller \inst{1}
\and E. Lagadec \inst{2}
\and J. H. Black \inst{1}
\and R. D. Oudmaijer \inst{3}
\and K. Justtanont \inst{1}
\and H. van Winckel \inst{4}
\and A. A. Zijlstra \inst{5}
}
\institute{Department of Earth and Space Sciences, Chalmers University of Technology, Onsala Space Observatory SE 439-92, Onsala, Sweden
\and Department of Astronomy, Cornell University, Ithaca, NY 14853-6801, USA
\and School of Physics and Astronomy, University of Leeds, Leeds, LS2 9JT, UK
\and Instituut voor Sterrenkunde, K.U. Leuven, Celestijnenlaan, 200D, 3001, Leuven, Belgium
\and Jodrell Bank Center For Astrophysics, The University of Manchester, Manchester, M13 9PL, UK
}

\date {Received  / Accepted}

\titlerunning{Investigating the nature of the Fried Egg nebula}
\authorrunning{S. H. J. Wallstr\"om et al.}

\abstract{Through CO mm-line and optical spectroscopy, we investigate
the properties of the Fried Egg nebula IRAS~17163--3907, which has recently been proposed to be one of the rare members of the yellow hypergiant class. The CO J=2--1 and J=3--2 emission arises from a region within 20$\arcsec$ of the star and is clearly associated with the circumstellar material. The CO lines show a multi-component asymmetrical profile, and an unexpected velocity gradient is resolved in the east-west direction, suggesting a bipolar outflow. This is in contrast with the apparent symmetry of the dust envelope as observed in the infrared.
The optical spectrum of IRAS~17163--3907 between 5100 and 9000~\r{A} was compared with that of the archetypal yellow hypergiant IRC+10420 and was found to be very similar. These results build on previous evidence that IRAS~17163--3907 is a yellow hypergiant. }


\maketitle

\section{Introduction}

Yellow hypergiants (YHGs) are a class of massive evolved stars. The YHG phase is thought to be short lived, and examples of such stars are rare, with only about a dozen known in our galaxy \citep{oud09}. However, YHG stars are characterised by intense mass-loss 
and can greatly impact the chemical enrichment of the interstellar medium, both with their massive winds and with their ultimate explosions as core-collapse supernovae.

According to \citet{deJ98}, the defining characteristics of YHGs are
a Ia$^{+}$ luminosity class, based on visible spectra, with H\,$\alpha$ emission in one or more broad components
and absorption lines that are broader than those of Ia supergiants of a similar spectral
type. The H\,$\alpha$ emission is a signature of an extended atmosphere and mass loss, while the comparatively
broad absorption lines signify large-scale photospheric motions such as pulsation.

These empirical characteristics have a physical interpretation, placing the YHGs in a
blueward evolutionary loop on the Hertzsprung-Russell diagram after the red supergiant phase. Only stars with
initial masses between 20 and 40~M$_{\odot}$ can take this path, and the timescale of the YHG phase is of the order of 100-1000 years, accounting for the rarity of this type of star. 
A YHG may subsequently
evolve through a phase as a luminous blue variable, and finally become a Wolf-Rayet star \citep{oud09}.

Among the most well-studied YHGs are HD~179821 (also known as IRAS~19114+0002 or AFGL~2343) and IRC+10420, both of which have prominent infrared excesses and dusty circumstellar envelopes, the latter having suffered a major ejection within the last 600 years \citep{hum97,hum02,oud96}. 
Very few YHG stars show resolved envelopes \citep{deJ98}, and studies of their mass-loss are limited (e.g. \citealp{cas07,din09}).

This paper focuses on the source IRAS~17163--3907 (also known
as Hen~3-1379, and hereafter referred to as IRAS~17163). It was discovered by \citet{hen76} and was first classified as a post-AGB star by \citet{leb89}. 
Using interstellar K\,I absorption in the optical spectrum, \citet{lag11b} placed a lower limit of 3.6~kpc on the distance to the star. An upper limit of $\sim$4.7~kpc was derived from the maximum visual extinction implied by diffuse interstallar band absorptions. Consequently, IRAS~17163 is at a distance four times larger than the distance of $\sim$1~kpc estimated by \citet{leb89}.
The revised luminosity of $\sim$5$\times$10$^5$~L$_{\odot}$, along with its position on a temperature-luminosity diagram close to the brightest YHGs like IRC+10420, means that IRAS~17163 is almost certainly a member of the yellow hypergiant class.

Despite being one of the brightest objects in the mid-IR sky, IRAS~17163 has so far been the subject of limited investigations. Mid-IR images of IRAS~17163 were obtained by Lagadec et al. (2011a;b) 
using the VLT-VISIR instrument. Two concentric spherical dusty shells were resolved within
2.5$\arcsec$ of the central star, with a warm ($\sim$200~K) dust mass of 0.04~M$_{\odot}$. The presence of two shells suggests an episodic enhancement of the mass-loss rate on timescales of a few hundred years.
IRAS~17163 was also imaged with \textit{Herschel} \citep{hut13}, revealing a symmetric dust shell $\sim$25$\arcsec$ in radius around the star. The mass of cool (60~K) dust in this shell is inferred to be as high as $\sim$0.17~M$_{\odot}$, implying a huge (although poorly constrained because of the unknown gas/dust ratio) circumstellar gas mass. However, dust imaging lacks kinematic information, preventing us from studying the mass loss and its timescale in more detail. In this paper, we present the first observations of CO rotational lines around IRAS~17163 and an optical spectrum of the star between 5100 and 9000~\r{A}.

\section{Observations} \label{Obs}

\subsection{APEX observations}

IRAS~17163 was observed with the Atacama Pathfinder EXperiment
(APEX) telescope\footnote{This publication is based on data acquired with the Atacama Pathfinder Experiment (APEX).
APEX is a collaboration between the Max-Planck-Institut fur Radioastronomie, the European Southern
Observatory, and the Onsala Space Observatory.} on April 9--12, 2014. A $\sim$100$\arcsec\times$100$\arcsec$ on-the-fly map in the CO J=2--1 and J=3--2 transitions was performed, centred on the star at R.A. 17:19:49.33, Dec. --39:10:37.9 (J2000). The data were taken with a scan spacing of 9$\arcsec$ in CO 2--1 (main beam size 27$\arcsec$) and 6$\arcsec$ in CO 3--2 (beam size 18$\arcsec$).

\begin{figure}[t] \begin{center}
\includegraphics[width=8cm]{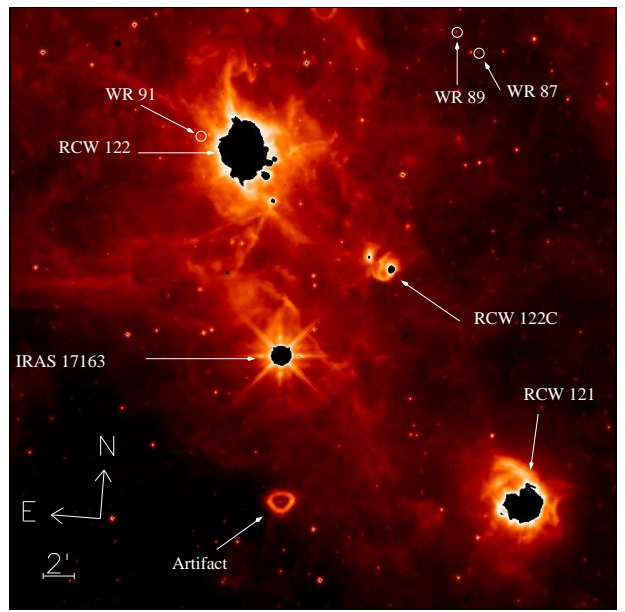}
\caption{WISE W3 band (11.56$\mu$m) image of the field around IRAS~17163, showing the nearby star-forming regions RCW 121 and 122.}
\label{fig:WISE}
\end{center} \end{figure}

The direction of IRAS~17163 lies close to the complex star-forming regions RCW~121 and RCW~122 (see Fig.~\ref{fig:WISE}),
which show bright CO emission at velocities between $v_{\mathrm LSR}=$ --40 and 0~km\,s$^{-1}$,
extended over a region of several square arcminutes \citep{arn08}. We adopted an on/off position-switching
calibration scheme, carefully selecting an off-position free from CO emission, $\sim$20$\arcmin$ 
south-east of the source.

The data were reduced within the standard single-dish
data reduction package CLASS\footnote{http://www.iram.fr/IRAMFR/GILDAS}.
The spectra have a total velocity coverage of 5720 and 3810~km\,s$^{-1}$ in CO 2--1 and 3--2, respectively. There are no line detections in either spectrum outside the velocity interval --40 to +110~km\,s$^{-1}$.

\begin{figure}[t] \begin{center}
\includegraphics[width=8.5cm]{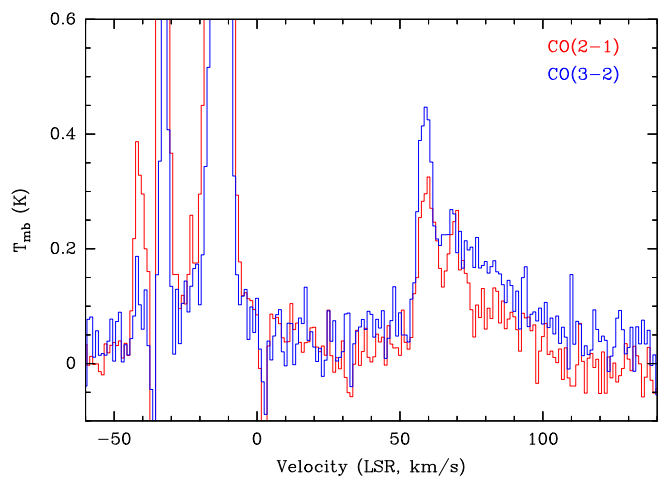}   
\caption{APEX spectra (velocity resolution 1~km\,s$^{-1}$) of the CO 2--1 (red) and 3--2 (blue) emission, averaged over the central 20$\arcsec$ around IRAS~17163. The emission between --40 and 0~km\,s$^{-1}$ is interstellar, while the emission between 50 and 110~km\,s$^{-1}$ is associated with the star.}
\label{fig:spec65}
\end{center} \end{figure}

\begin{figure}[ht]
\centering
\includegraphics[width=8cm]{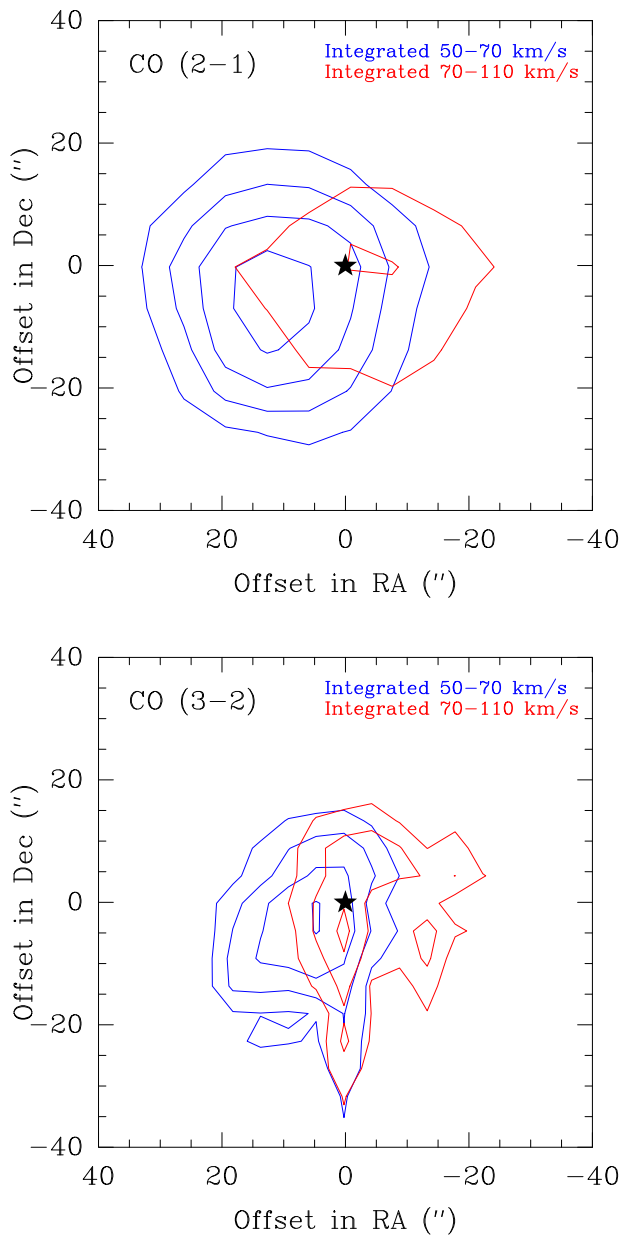} 
\caption{\label{contours} Contours of the integrated CO 2--1 and 3--2 emission between 50--70~km\,s$^{-1}$ (blue) and 70--110~km\,s$^{-1}$ (red). All contours start at 5$\sigma$ and are spaced by 5$\sigma$ (1$\sigma$=0.27~K\,km\,s$^{-1}$ for CO 2--1 and 0.33~K\,km\,s$^{-1}$ for CO 3--2). The main beam sizes are 27$''$ for CO 2--1 and 18$''$ for CO 3--2. The star at (0,0) marks the position of IRAS~17163. }
\end{figure}

The components between --40 and 0~km\,s$^{-1}$ correspond to CO emission from the RCW~121 and RCW~122 complexes
\citep{arn08}. An additional emission feature is seen between $v_{\mathrm LSR}$=50--110~km\,s$^{-1}$ (Fig.~\ref{fig:spec65}).
In contrast to the widespread CO features in the velocity range --40 to 0~km\,s$^{-1}$, this component is confined to within $\sim$20$\arcsec$ of IRAS~17163 (Fig.~\ref{contours}, \ref{fig:5x5}) and is clearly associated with the circumstellar envelope. The CO radial velocity, $\sim$70~km\,s$^{-1}$, is also completely incompatible with galactic rotation at any distance towards IRAS~17163.

\subsection{Optical spectrum from Mercator}

IRAS~17163 was observed on August 10, 2009, with the fibre-fed spectrograph HERMES \citep{ras11} attached to the 1.2-meter Mercator telescope at Roque de los Muchachos observatory, Spain. The fibre has an aperture of 2.5$\arcsec$ on the sky. Four exposures of 20 minutes each were taken in concatenation, and the data were reduced with the dedicated pipeline. The resulting flat-fielded, merged, and
wavelength-calibrated spectrum was normalized by the continuum level and is presented in Appendix~\ref{app:spectra}.
Note that the spectrum was not corrected for telluric features. The spectral resolution is $\sim$4~km\,s$^{-1}$ and the signal-to-noise ratio ranges between $\sim$5--100 depending on the wavelength, with lower values in the blue due to the severe reddening of the star.

\section{Results} \label{Results}

\subsection{CO emission} \label{sect:COprofile}

The CO J=2--1 and J=3--2 emission lines are broad ($\sim$60~km\,s$^{-1}$) and asymmetrical, with peaks between $v_{LSR}$=50--70~km\,s$^{-1}$ and a fainter plateau between 70--110~km\,s$^{-1}$ (Fig.~\ref{fig:spec65}). Integrating over these blue (50--70~km\,s$^{-1}$) and red (70--110~km\,s$^{-1}$) emission components shows that the CO emission is concentrated within $\sim$20$\arcsec$ of the star and that there is a clear spatial separation between the two components (Fig.~\ref{contours}). In CO 2--1 the blue and red components' emission peaks are offset from the star by $\sim$14$\arcsec$ and 4$\arcsec$, respectively, giving a spatial separation of $\sim$18$\arcsec$, in a mainly east-west direction. This velocity gradient is resolved within the APEX beam and suggests an asymmetrical outflow with a velocity of $\sim$30~km\,s$^{-1}$. 
The velocity gradient is less apparent in CO 3--2, partly because the red component has a low signal-to-noise ratio. In addition, the CO 3--2 emission is generally concentrated closer to the star, with the blue and red components' emission peaks offset from the star by $\sim$6$\arcsec$ east and 5$\arcsec$ south, respectively. This may be because the gas closer to the star is warmer and hence brighter in CO 3--2 than 2--1. The CO (3--2)/(2--1) peak intensity ratio is $\sim$1.4 at the star, and decreases to $\sim$1.2 at (+10$\arcsec$,0).

\begin{figure*}[t] \begin{center}
\includegraphics[width=14cm]{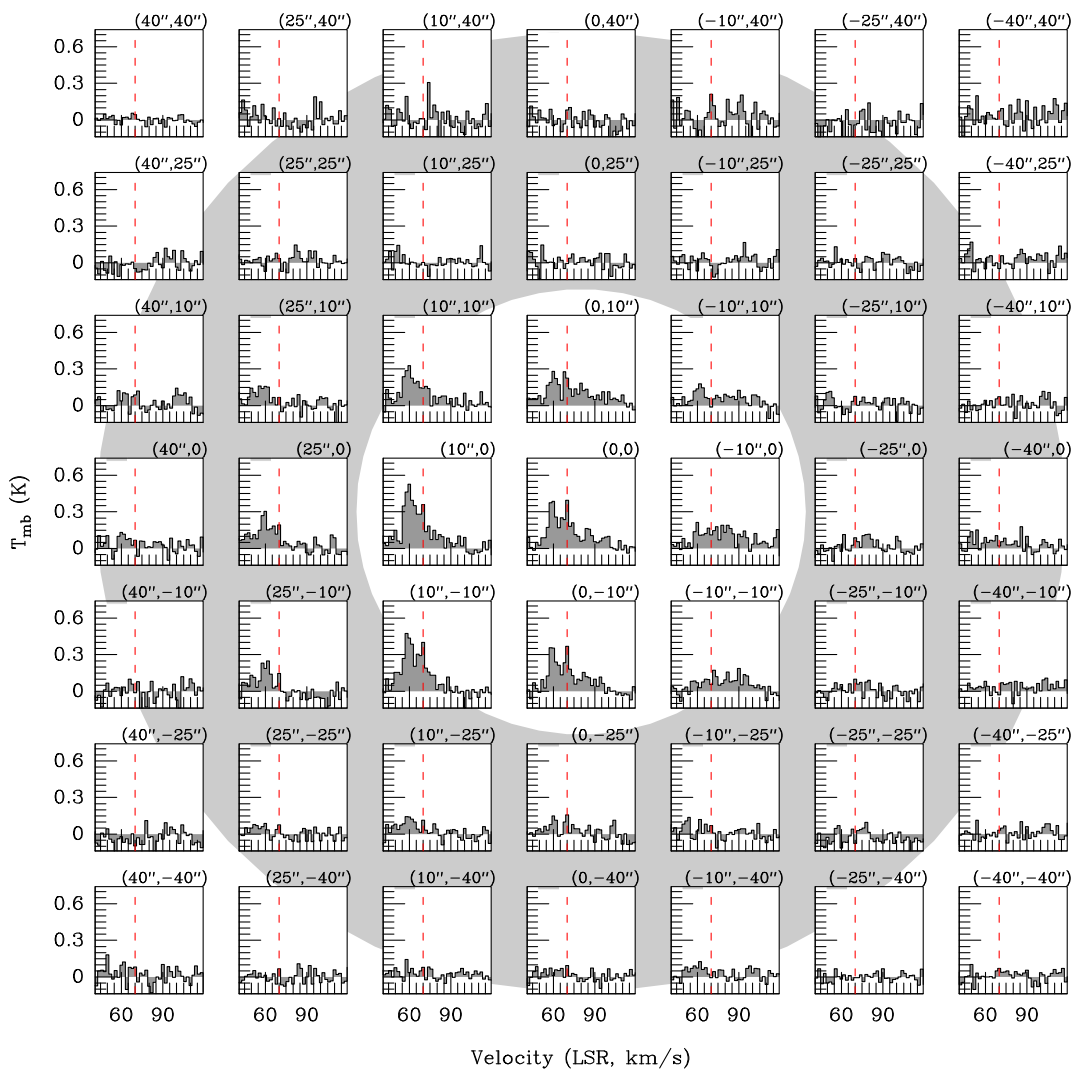} 
\caption{Map of the APEX CO 2--1 emission (velocity resolution 2~km\,s$^{-1}$) around IRAS~17163, with the star at (0,0). The grey background ring shows the location and extent of the 25$\arcsec$ dust ring. Most of the CO emission is found inside this dust ring. Note also the dashed red lines that mark 70~km\,s$^{-1}$, dividing the blue and red components of the emission profile. These components are spatially separated, with the blue component peaking around (10$\arcsec$,0) and the red component around (--10$\arcsec$,0). The equivalent map for CO 3--2 is shown in Appendix~\ref{app:32}.}
\label{fig:5x5}
\end{center} \end{figure*}

Previous infrared observations have found a warm dust mass of 0.04~M$_\odot$ in shells within 2.5$\arcsec$ of the star \citep{lag11b}, and a cooler dust ring peaking $\sim$25$\arcsec$ from the star (extending in radius from 18$\arcsec$ to 40$\arcsec$), with a dust mass of 0.17~M$_\odot$ \citep{hut13}. This large amount of dust implies a significant gas mass, especially in the 25$\arcsec$ ring, but the CO observations show little emission in this area (Fig.~\ref{fig:5x5}). The CO emission is found mainly within $\sim$20$\arcsec$ of the star, which is most likely due to 
the CO farther away being photodissociated by the interstellar UV radiation field. 
The asymmetry of the CO emission contrasts with the apparent symmetry of the dust rings at both $\sim$2.5$\arcsec$ and $\sim$25$\arcsec$ from the star. However, we note that a slight asymmetry can be seen in the \citet{hut13} dust image, along the same direction as the CO velocity gradient.

The asymmetry of the CO line profile and the velocity gradient mean we cannot model the mass loss as a spherically symmetric wind. This precludes us from reliably estimating the mass-loss rate. However, catalogues of AGB and post-AGB stars indicate that a CO envelope of $\sim$20$\arcsec$ would require a large sustained mass-loss rate of around 10$^{-5}$~M$_{\odot}$\,yr$^{-1}$ \citep{lou93}. This is in the range of expected mass-loss rates for YHGs.

To give a rough estimate of the circumstellar gas mass, we consider that if the integrated CO line intensity (5.85 K\,km/s) were that of a gravitationally bound spherical interstellar cloud of 20$\arcsec$ radius at a distance of 4 kpc, then the application of a standard CO/H$_2$ conversion factor of 2$\times 10^{20}$~cm$^{-2}$/(K\,km\,s$^{-1}$) would imply a gas mass $\sim$6~M$_{\odot}$. By coincidence, this is similar to the $\sim$7~M$_\odot$ mass of the 25$\arcsec$ dust ring derived by \citet{hut13}, who assumed a gas-to-dust mass ratio of 40.

\subsection{Optical spectroscopy} \label{sect:optical}

\begin{figure}[ht] \begin{center}
\includegraphics[width=7cm]{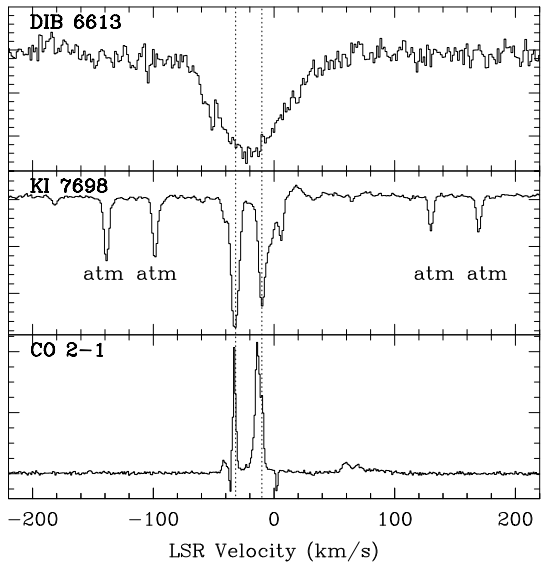} 
\caption{Interstellar absorption lines from the optical spectrum, with wavelengths in \r{A}, at the same velocity as the interstellar CO emission from APEX observations. The dotted vertical lines are at --32 and --10~km\,s$^{-1}$.}
\label{fig:ISlines}
\end{center} \end{figure}

The optical spectrum of IRAS~17163 is rich in lines, mainly in emission, but also some in absorption. Of the identified lines, the largest number correspond to Fe II, Cr II, and Ca II in emission and N I and H I in absorption. Interstellar absorption lines of K\,I 
and diffuse interstellar bands (DIBs) have velocities corresponding to the CO emission from the surrounding star-forming regions (see Fig.~\ref{fig:ISlines}).

\begin{figure}[ht] \begin{center}
\includegraphics[width=8cm]{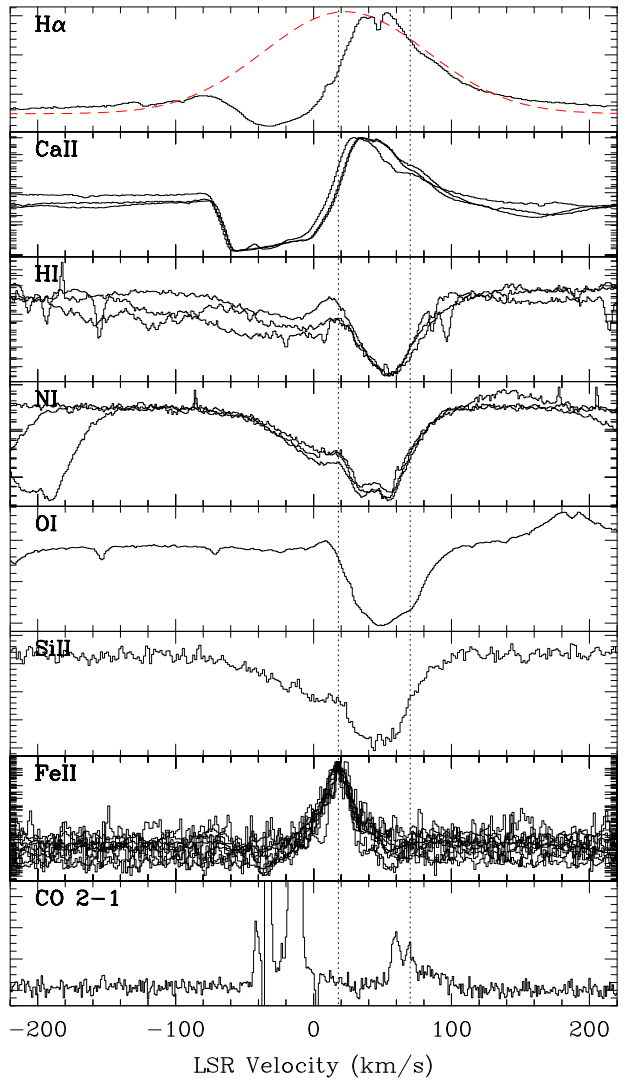} 
\caption{Various lines from the optical spectrum. A Gaussian fit to the central part of the H\,$\alpha$ line (FWHM=140~km\,s$^{-1}$) is shown in dashed red. The dotted vertical lines are at 18 and 70~km\,s$^{-1}$, matching the Fe\,II and centre of the CO emission, respectively. The CO emission at negative velocities is interstellar.}
\label{fig:optlines}
\end{center} \end{figure}

The brightest line, H\,$\alpha$ (at 6563 \r{A}), has a P-Cygni profile with a FWHM of $\sim$140~km\,s$^{-1}$ (see Fig.~\ref{fig:optlines}) and broad wings extending out to a width of $\sim$2400~km\,s$^{-1}$.
A P-Cygni line profile is indicative of a substantial outflow of material, and the broad H\,$\alpha$ wings may be due to scattering by free electrons, as suggested by \citet{hum02} for IRC+10420. Three Ca\,II lines (at 8498, 8542, and 8662 \r{A}) also show P-Cygni profiles, with line widths similar to that of H\,$\alpha$. A relatively large number ($\sim$50) of lines remains unidentified in the spectrum.

The central velocities of various lines range from +10 to +43~km\,s$^{-1}$ (see Fig.~\ref{fig:optlines}), 
but both H\,$\alpha$ and the narrow (FWHM=25~km\,s$^{-1}$) Fe\,II lines are centred on a velocity of +18~km\,s$^{-1}$. Given their excitation, the Fe\,II lines most likely arise near the stellar photosphere and are indicative of the stellar velocity. 
However, their velocity is blueshifted by about 50~km\,s$^{-1}$ relative to the velocity of the CO emission. Possible explanations for this are discussed in\ Sect. \ref{sect:velocity}.

The absolute magnitude ($M_{{\rm V}}$) of IRAS~17163 can be estimated from the equivalent width ($W_{\lambda}$) of the O\,I\,(7774) triplet, which has been shown to be a good indicator of $M_{{\rm V}}$ for stars with spectral types A to G (\citealt{are03}, and references therein). We measure $W_{\lambda}$(O\,I\,(7774))=3.14$\pm$0.01 \r{A}, which is just outside the range
of the $W_{\lambda}$(O\,I)--$M_{{\rm V}}$ calibration by \citet{are03}. Nevertheless, a straight extrapolation suggests an $M_{{\rm V}}$ close to --10, comparable with the values similarly derived for the two YHGs IRC+10420 and HD~179821 \citep{oud09}.

\section{Discussion} \label{Discussion}

\subsection{Velocity discrepancy} \label{sect:velocity}

IRAS~17163 shows broad CO emission, centred around 70~km\,s$^{-1}$, with a velocity gradient across the circumstellar envelope. We hence take this to be the systemic velocity of the star. The star also shows multiple optical Fe\,II lines, centred on a velocity of 18~km\,s$^{-1}$. This should also be indicative of the systemic velocity, yet it is blueshifted by $\sim$50~km\,s$^{-1}$. 

A velocity discrepancy between CO and optical line velocities is also found in the archetypal YHG IRC+10420, although it is smaller. In this object, the optical emission lines are blueshifted by 15$-$25~km\,s$^{-1}$ compared with the CO systemic velocity at +75~km\,s$^{-1}$. However, the optical lines still fall within the CO velocity range. In the YHG HD~179821, the CO and optical lines coincide at +100~km\,s$^{-1}$.

There are several possible explanations for the large velocity discrepancy in IRAS~17163. First is photospheric pulsations: yellow hypergiant stars are characterised by large-scale photospheric variability. The YHG $\rho$~Cas has shown radial velocity variations of up to 35~km\,s$^{-1}$ associated with outburst activity, on a timescale of $\sim$100 days \citep{lob03}. Furthermore, models show that YHG pulsational variability can be up to 100~km\,s$^{-1}$ \citep{fad11}. As we only have one epoch of optical observations of IRAS~17163, we cannot constrain its photospheric variability, so it is possible that our optical spectrum was taken near the peak of a pulsational cycle when the photospheric lines were blueshifted by $\sim$50~km\,s$^{-1}$.

Another possibility is a close binary companion. In this case, the CO emission would indicate the systemic velocity of the binary system, while the Fe\,II lines arise only from our YHG. IRAS~17163 would need to have an orbital velocity of at least 50~km\,s$^{-1}$ to explain the velocity discrepancy between the CO and Fe\,II lines. For example, the LBV binary MWC~314 has an orbital speed of $\sim$100~km\,s$^{-1}$ \citep{lob13}. The binary hypothesis also requires that the companion be relatively optically faint as there is no indication of lines at a very different velocity in the optical spectrum. The companion would have to be $\sim$100 (the highest signal-to-noise ratio in the spectrum) times less luminous than IRAS~17163, i.e., $\sim$5$\times$10$^3$~L$_{\odot}$. For a main sequence companion, this rules out only the heaviest (greater than $\sim$10~M$_\odot$) stars.

In summary, the radial velocity of the Fe\,II optical lines is compatible with the systemic velocity derived from the CO observations. We note that a systemic velocity of $\sim$70~km\,s$^{-1}$ means that the star has a very high peculiar velocity, redshifted by about 100~km\,s$^{-1}$ compared with galactic rotation at 4 kpc \citep{rei09}. However, about 20\% of massive stars in the Milky Way are runaways (e.g. \citealp{mas98}), so this is not implausible.

\subsection{Nature of IRAS~17163}

\subsubsection{Distance} 

The knowledge of the distance is a key parameter for defining the nature and physical properties of IRAS~17163, as it determines its absolute luminosity and hence its position in the Hertzsprung--Russell diagram.
The signature of foreground interstellar clouds in its optical spectrum
allowed \citet{lag11b} to conclude that the distance to IRAS~17163 is at least 3.6 kpc. 
This large distance is also supported by recent trigonometric parallax measurements of the star-forming region RCW~122, placing it at a distance of 3.38$\pm$0.3~kpc \citep{wu12}. RCW~122 is part of the complex whose CO emission matches the K\,I absorption in our spectra (Fig.~\ref{fig:ISlines}), and is hence in front of IRAS~17163. 

The knowledge of the absolute magnitude, $M_{\rm V}$ (see $\S$\,\ref{sect:optical}), of IRAS~17163 provides an alternative method of determining its distance, $D$ (in pc), using the equation
\begin{equation}
m_{\rm V} - M_{\rm V} - A_{\rm V} + 5 = 5 \times \log{D}
\label{eqn:mag}
,\end{equation}
where $m_{\rm V}$ is the apparent visual magnitude and $A_{\rm V}$ the visual extinction along the line of sight to the star.
We take $m_{\rm V}$=13.03 mag and $E(B-V)$=4.10 mag from \citet{leb89} and assume that there has been no significant variation in the star between the date of the photometric observations (1988) and the more recent spectroscopic observations of O\,I (2009), from which $M_{\rm V} \sim -10$ is determined.
A standard conversion between visual extinction and reddening of $A_{\rm V}=3.1 \times E(B-V)$ is used.
Taking the photometric reddening at face value, Eq.~(\ref{eqn:mag}) gives a distance to IRAS~17163 of $\sim$1~kpc. Alternatively, an estimate of the interstellar reddening can be found from the depths of DIBs in the optical spectrum. Taking the highest interstellar reddening value derived by \citet{lag11b}, we calculate a distance of $\sim$7~kpc.
The uncertainties in the amount and properties of interstellar and circumstellar dust limit the application of this method for constraining the distance to IRAS~17163. In any case, the distance interval ($\sim$1--7~kpc) is consistent with, but less restrictive than, that the 3.6$-$4.7~kpc found by \citet{lag11b}. Throughout this paper we have assumed a distance of 4~kpc.

\subsubsection{Mass of the star}

The terminal velocities of stellar winds are typically of the order of the stellar escape velocity (see e.g. \citealp{abb78,lam99}), which we can use to roughly estimate the current stellar mass. 
We can estimate the wind terminal velocity from the Ca\,II P Cygni profile, where the blue edge of the absorption represents the highest wind speed: --80~km\,s$^{-1}$. As the wind originates at the systemic velocity, 70~km\,s$^{-1}$, this gives a terminal velocity of 150~km\,s$^{-1}$. 

From the luminosity of 5$\times10^5$~L$_\odot$ and an assumed effective temperature of 8000~K (from the determination of IRAS~17163 as a late B- or early A-type star by \citet{lag11b}), we calculate a stellar radius of $\sim$370~R$_\odot$. Then, equating the terminal velocity of the stellar wind, 150~km\,s$^{-1}$, with the escape velocity $v_e=\sqrt{2GM_*/R_*}$ , we derive a stellar mass of $\sim$22~M$_\odot$. Summing this value with the estimate of the circumstellar mass (see Sect. \ref{sect:COprofile}) yields an initial mass  of about 30~M$_\odot$, consistent with the expected YHG initial mass range of 20 to 40~M$_\odot$.

\subsubsection{Comparison with IRC+10420}

There is a wealth of similarity between IRAS~17163 and IRC+10420, the archetypal YHG. Their close proximity in the Hertzsprung--Russell diagram prompted the classification of IRAS~17163 as a YHG, and both objects have absolute magnitudes close to $M_{\rm V}$=--10 (see Sect. \ref{sect:optical}). Both objects show distinct circumstellar shells, with ejection timescales of a few hundred years \citep{lag11b,oud09}. 

Furthermore, the optical spectrum of IRAS~17163 closely resembles the peculiar optical spectrum of IRC+10420 \citep{oud98}. The two spectra are shown, overlaid, in Appendix~\ref{app:spectra}.
Both objects show strong H\,$\alpha$ and Ca\,II lines. In IRAS~17163 these lines have clear P-Cygni profiles, indicative of circumstellar outflow. In IRC+10420 they show strong central absorptions.
There are a few other differences between the two spectra: 
IRAS~17163 is somewhat more line rich, with about a dozen lines that are not seen at all in the spectrum of IRC+10420. Conversely, there are two Fe\,I lines seen only in the IRC+10420 spectrum. The Mg\,II lines around 7890 \r{A} are seen in emission in IRAS~17163 and in absorption in IRC+10420.

Overall, the optical spectra of IRAS~17163 and IRC+10420 are very similar. This, along with the other similarities between these two objects, supports the classification of IRAS~17163 as a yellow hypergiant.

\section{Conclusions} \label{Conclusion}

IRAS~17163 was recently proposed to be a member of the rare class of yellow hypergiants by \citet{lag11b} based on its location on a Hertzsprung--Russell diagram close to the archetypal yellow hypergiant IRC+10420.
To further investigate the nature of the star, we obtained CO J=2--1 and J=3--2 APEX observations and a high-resolution Mercator spectrum between 5100 and 9000 \r{A}.
Our findings can be summarized as follows:

\begin{itemize}
\item We observe CO emission associated with IRAS~17163, at a radial velocity around $v_{\mathrm LSR}=$+70~km\,s$^{-1}$.
\item The CO line profile is broad ($\Delta v \sim$60~km\,s$^{-1}$) and asymmetric, with a blue component showing multiple peaks and a fainter, broader red component.
\item A velocity gradient is resolved across the circumstellar envelope: a $\sim$18$\arcsec$ offset between the blue and red components' emission peaks, suggesting an asymmetrical outflow in an approximately east-west direction.
\item There is a velocity discrepancy between the CO emission, at $\sim$70~km\,s$^{-1}$, and the optical Fe\,II lines at 18~km\,s$^{-1}$. This might be explained by large photospheric pulsations in the star or by a binary companion. 
\item The optical spectrum is line rich and very similar to the peculiar spectrum of the archetypal yellow hypergiant star IRC+10420.
\item The P-Cygni profiles of the H\,$\alpha$ and Ca\,II lines signify a substantial amount of outflowing material.
\item The absolute magnitude was estimated using the equivalent width of the O\,I (7774 \r{A}) triplet to be $M_{\rm V}$=--10, comparable to that of IRC+10420.
\end{itemize}

We conclude that the similarity in properties between IRAS~17163 and IRC+10420 reinforces the classification of IRAS~17163 as a yellow hypergiant. The CO emission from this object is complex and has raised questions about how it fits in with previous observations. Further observations of the circumstellar envelope at high angular resolution, for instance, with ALMA, will be required to reveal the distribution of the molecular gas and further study the mass-loss activity.

\begin{acknowledgement}

We would like to thank the director of the Onsala Space Observatory, Hans Olofsson, for generous allocation of DDT time. We also thank Per Bergman for his help with our APEX observations and for helpful discussions.

This publication makes use of data products from the Wide-field Infrared Survey Explorer, which is a joint project of the University of California, Los Angeles, and the Jet Propulsion Laboratory/California Institute of Technology, funded by the National Aeronautics and Space Administration.

\end{acknowledgement}

\onecolumn 

\appendix

\section{Map of CO 3--2 emission} \label{app:32}

\begin{figure*}[h] \begin{center}
\includegraphics[width=14cm]{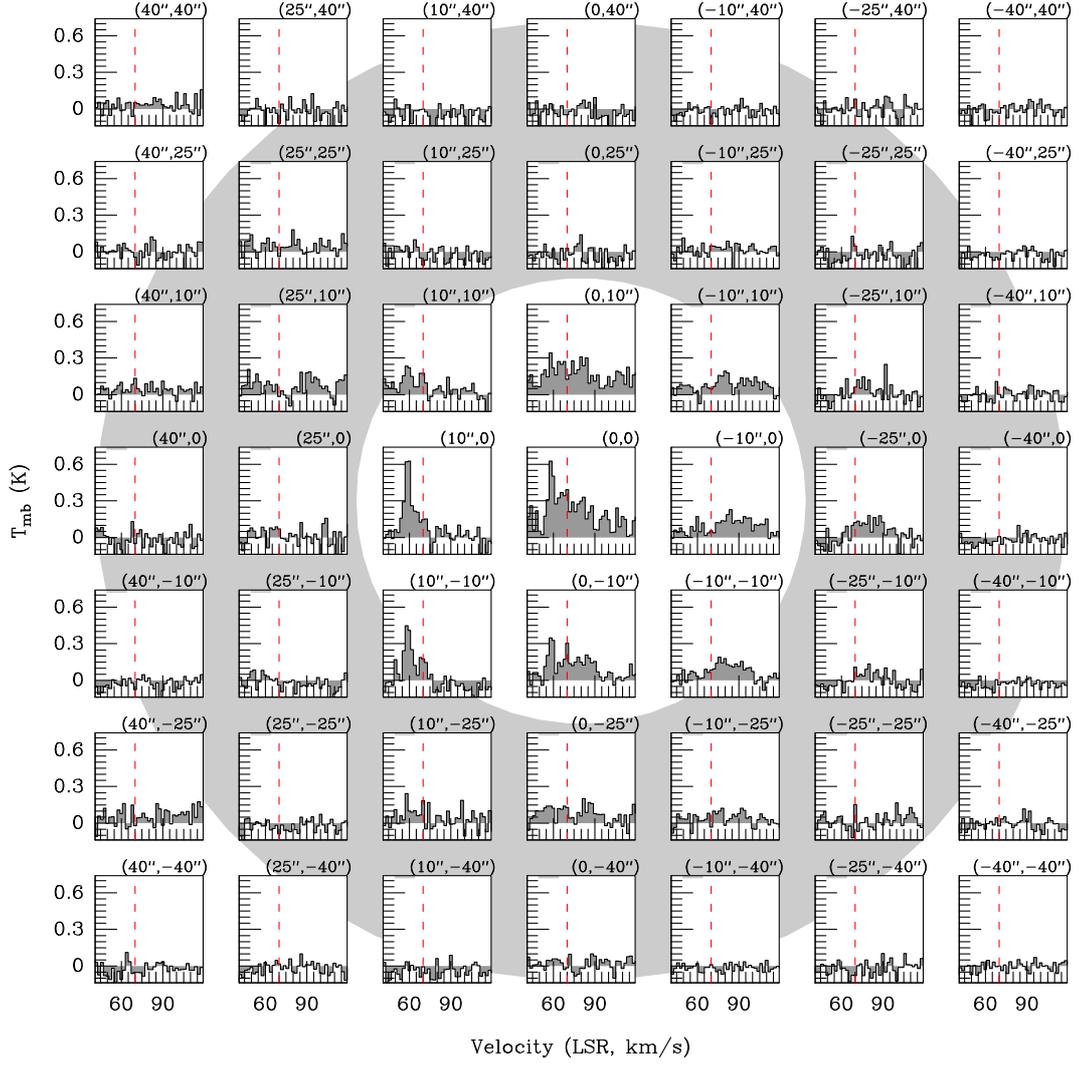} 
\caption{Map of the APEX CO 3--2 emission (velocity resolution 2~km\,s$^{-1}$) around IRAS~17163, with the star at (0,0). The grey background ring shows the location and extent of the 25$\arcsec$ dust ring. Most of the CO emission is found inside this dust ring. Note also the dashed red lines that mark 70~km\,s$^{-1}$, dividing the blue and red components of the emission profile. These components are spatially separated, with the blue component peaking around (10$\arcsec$,0) and the red component around (--10$\arcsec$,0).}
\label{fig:32_7x7}
\end{center} \end{figure*}

\newpage

\section{Optical spectrum of IRAS~17163, overlaid with the spectrum of IRC+10420} \label{app:spectra}

\begin{figure*}[!ht] \begin{center}
\includegraphics[width=0.9\textwidth]{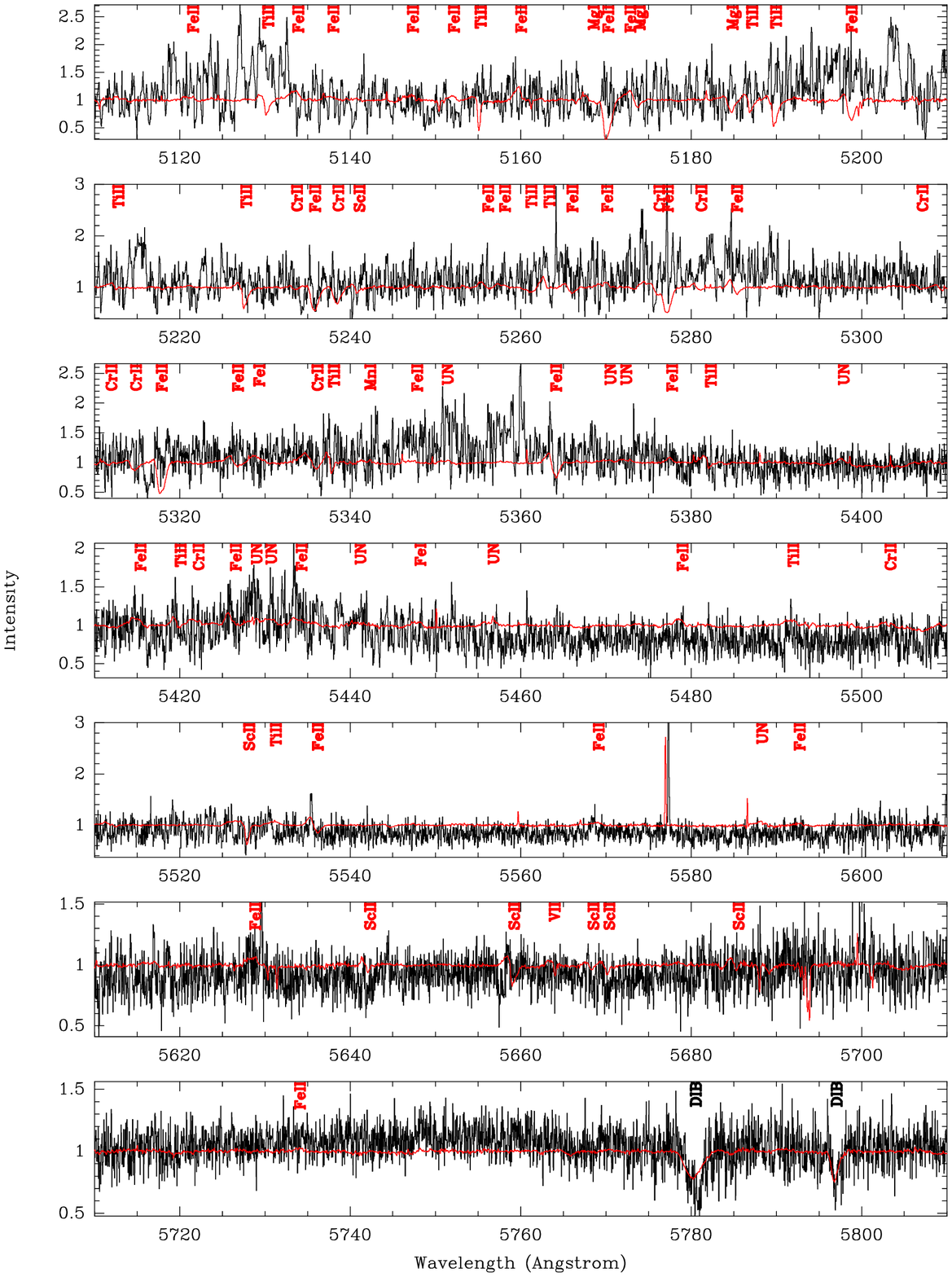}
\end{center} \end{figure*}

\begin{figure*}[htbp] \begin{center}
\includegraphics[width=\textwidth]{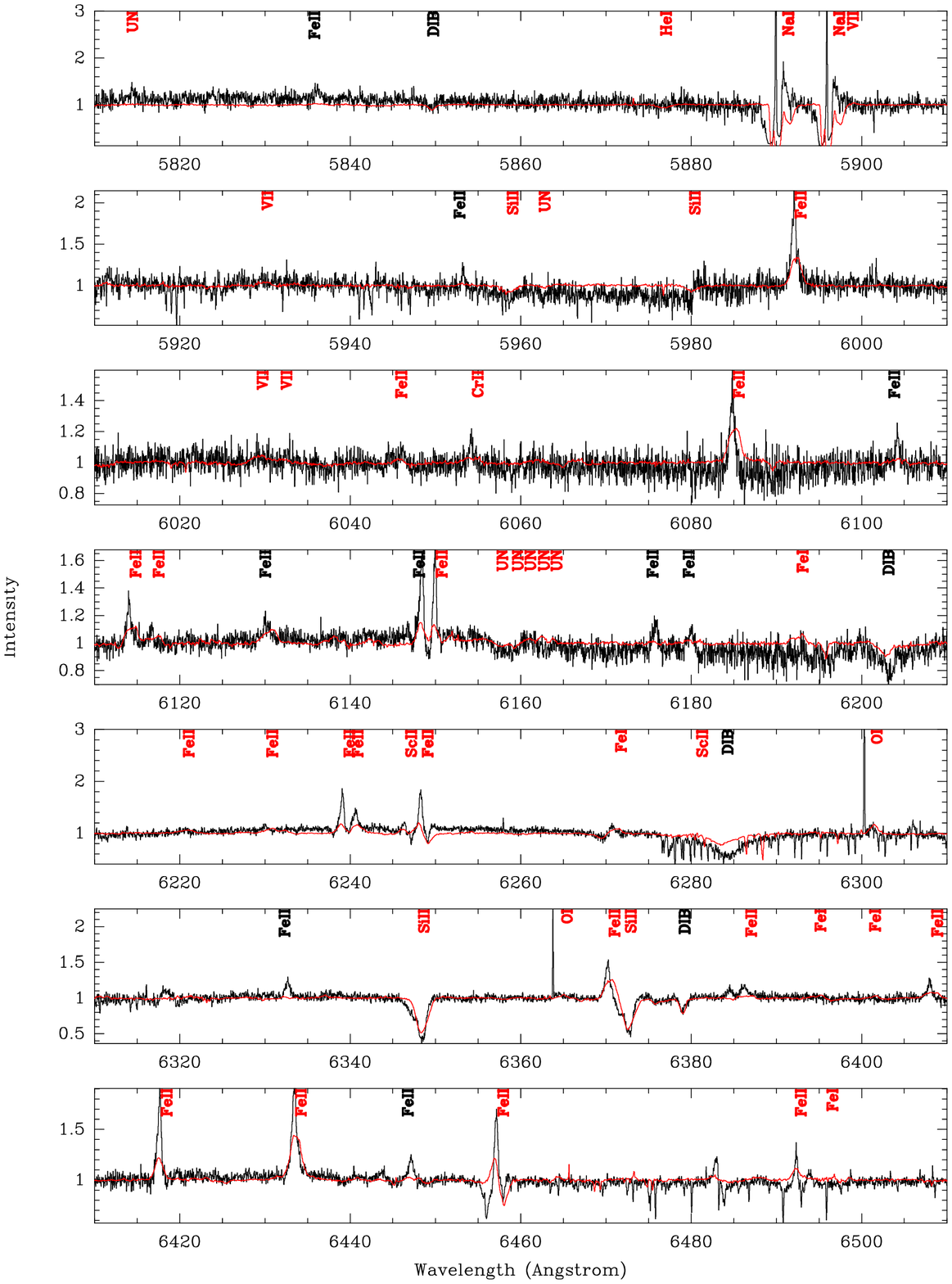}
\end{center} \end{figure*}

\begin{figure*}[htbp] \begin{center}
\includegraphics[width=\textwidth]{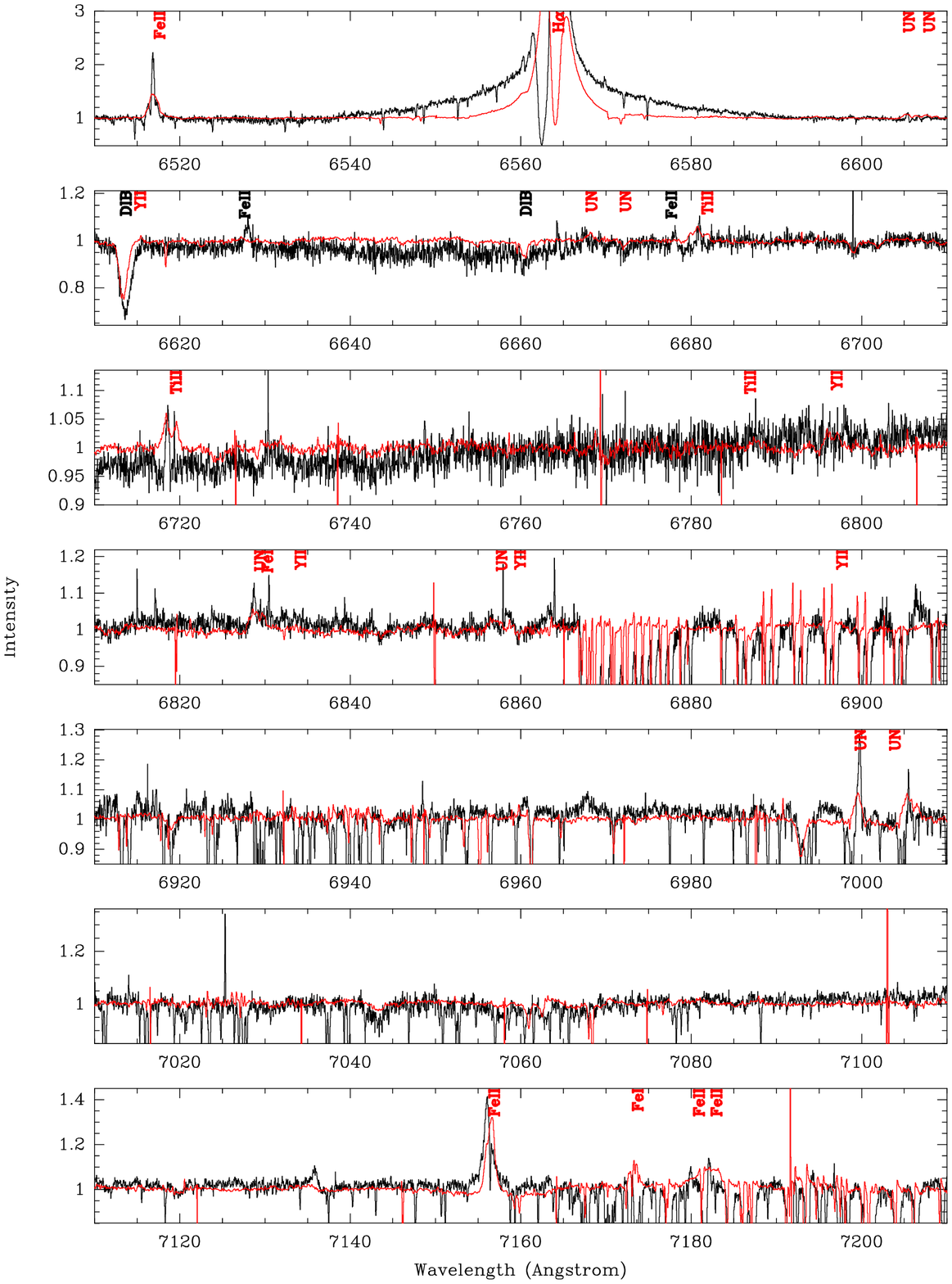}
\end{center} \end{figure*}

\begin{figure*}[htbp] \begin{center}
\includegraphics[width=\textwidth]{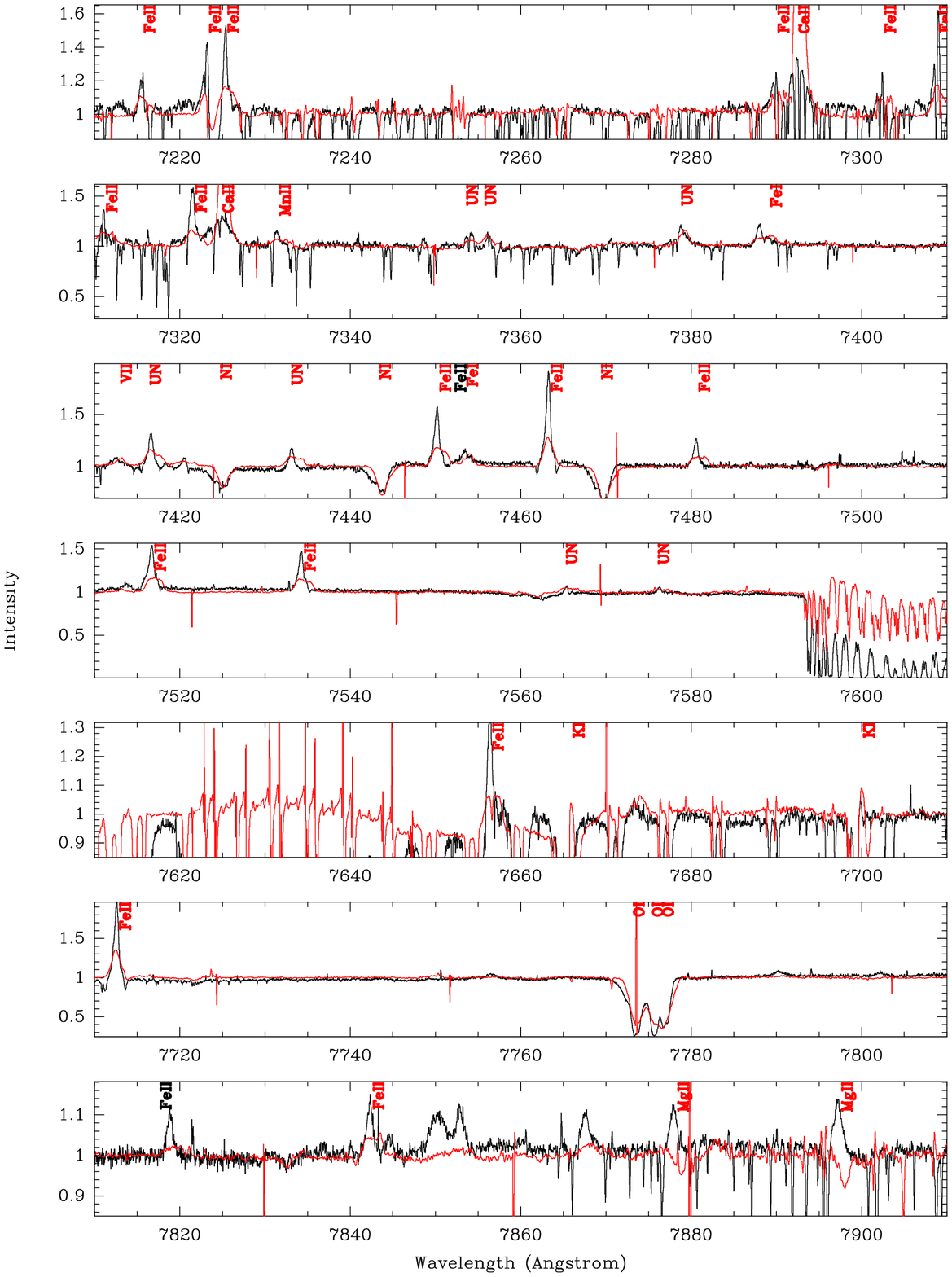}
\end{center} \end{figure*}

\begin{figure*}[htbp] \begin{center}
\includegraphics[width=\textwidth]{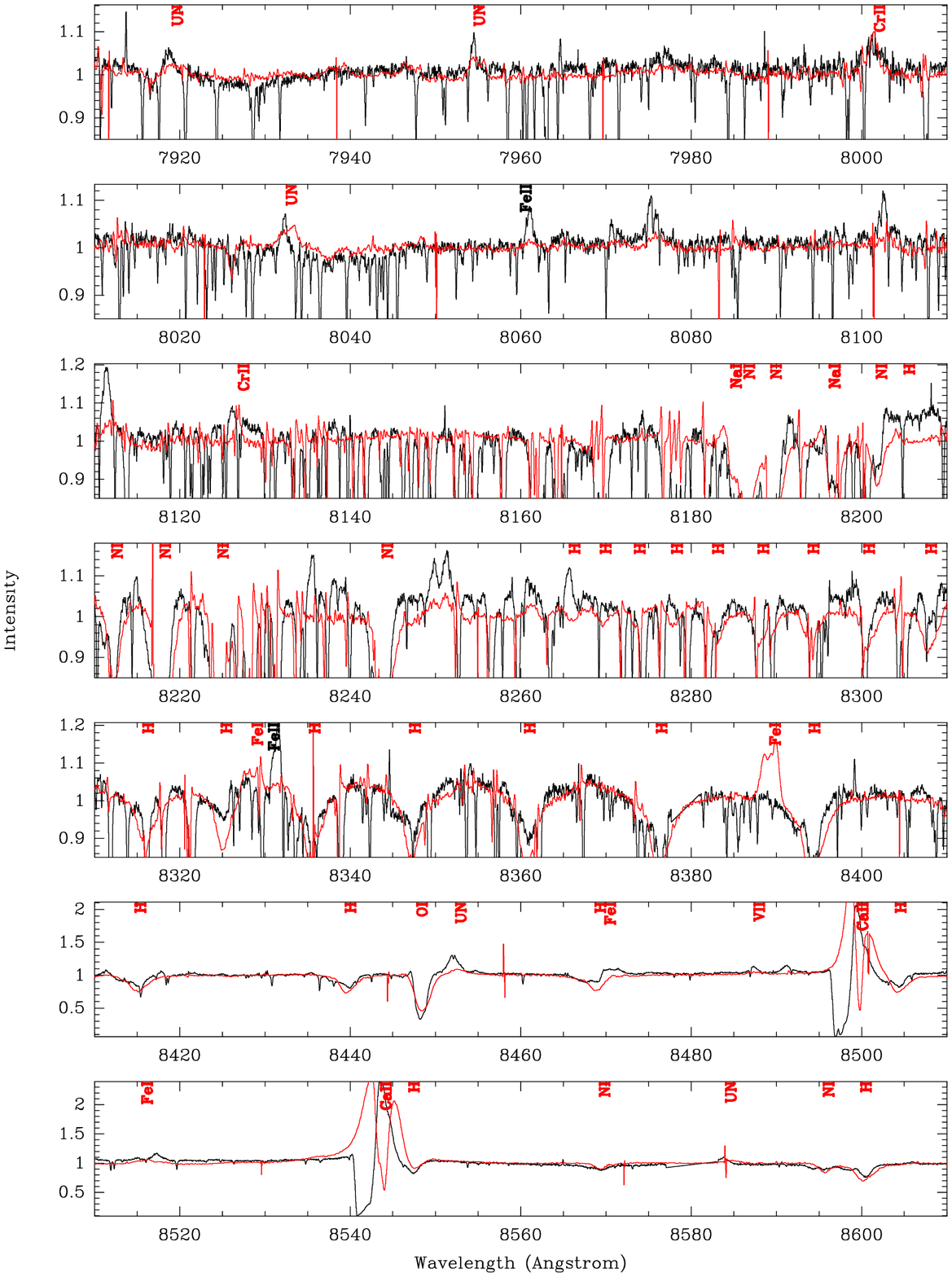}
\end{center} \end{figure*}

\begin{figure*}[htbp] \begin{center}
\includegraphics[width=\textwidth]{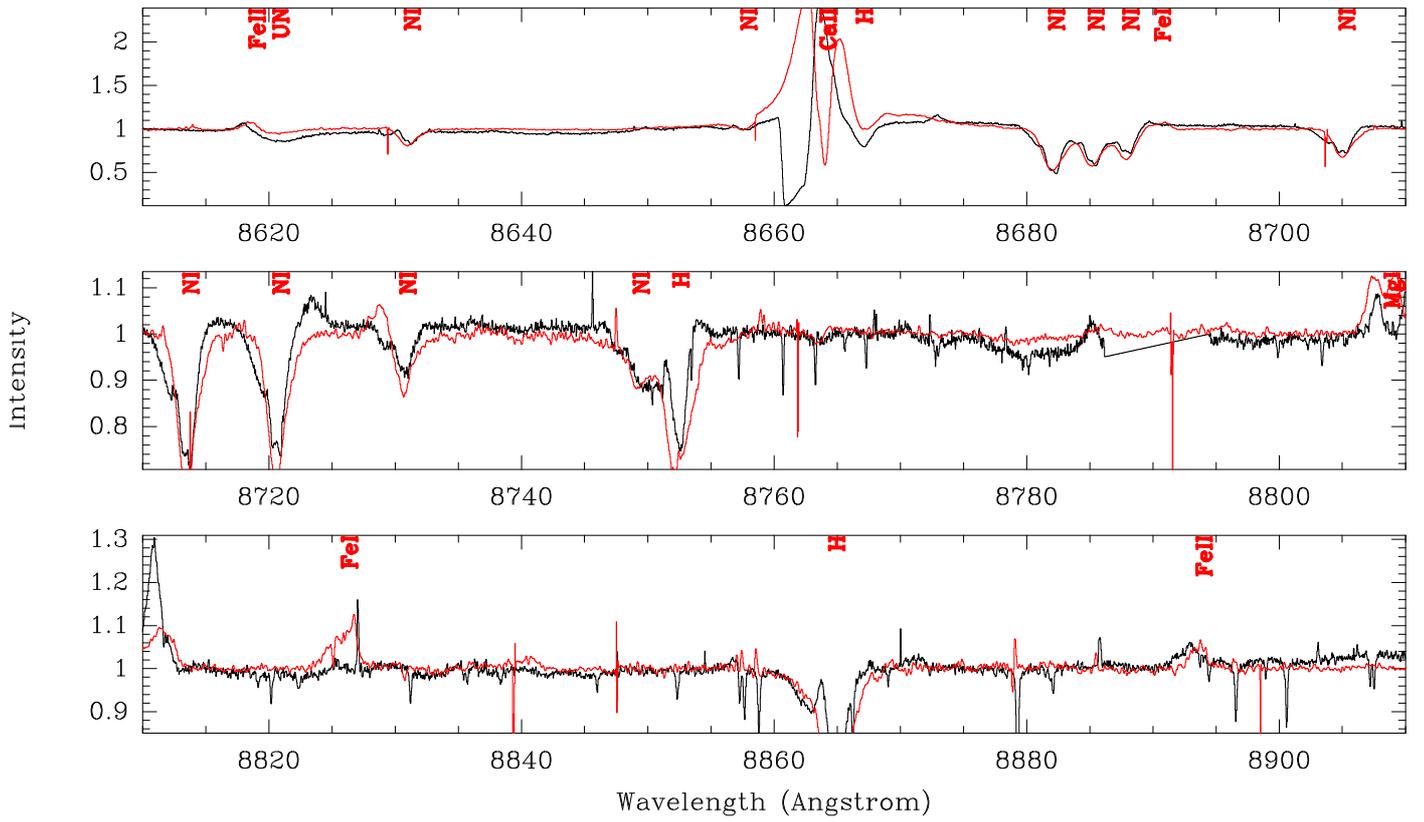}
\caption{Optical spectrum of IRAS~17163, in black, overlaid with the spectrum of IRC+10420 (from \citealp{oud98}), in red. The line identifications in red were made for the IRC+10420 spectrum by \citet{oud98} (with UN indicating an unidentified line, and H hydrogen recombination lines), and have been corrected for a systemic velocity of 75~km\,s$^{-1}$. The Fe\,II identifications in black have been corrected for a line velocity of 18~km\,s$^{-1}$.}
\end{center} \end{figure*}


\begin{thebibliography}{}


\bibitem[Abbott(1978)]{abb78} Abbott, D. C. 1978, \apj, 225, 893
\bibitem[Arellano Ferro et al.(2003)]{are03} Arellano Ferro, A.; Giridhar, S. \& Rojo Arellano, E.; et al. 2003, RMxAA, 39, 3
\bibitem[Arnal et al.(2008)]{arn08} Arnal, E. M.; Duronea, N. U. \& Testori, J. C. 2008, \aap, 486, 807
\bibitem[Bergman et al.(2011)]{ber11} Bergman, P.; Parise, B.; Liseau, R.; et al. 2011, \aap, 531, 8
\bibitem[Castro-Carrizo et al.(2007)]{cas07} Castro-Carrizo, A.; Quintana-Lacaci, G.; Bujarrabal, V.; et al. 2007, \aap, 465, 457
\bibitem[Dinh-V.-Trung et al.(2009)]{din09} Dinh-V.-Trung; Muller, S.; Lim, J.; et al. 2009, \apj, 697, 409
\bibitem[de Jager(1998)]{deJ98} de Jager, C. 1998, \aapr, 8, 145
\bibitem[Fadeyev(2011)]{fad11} Fadeyev, Y. A. 2011, Astronomy Letters, 37, 403
\bibitem[Henize(1976)]{hen76} Henize, K. G. 1976, ApJS, 30, 491
\bibitem[Humphreys et al.(1997)]{hum97} Humphreys, R. M.; Smith, N.; Davidson, K.; et al. 1997, \aj, 114, 2778
\bibitem[Humphreys et al.(2002)]{hum02} Humphreys, R. M.; Davidson, K. \& Smith, N. 2002, \aj, 124, 1026
\bibitem[Hutsem\'ekers et al.(2013)]{hut13} Hutsem\'ekers, D.; Cox, N. L. J. \& Vamvatira-Nakou, C. 2013, \aap, 552, 6
\bibitem[Lagadec et al.(2011b)]{lag11b} Lagadec, E.; Zijlstra, A. A.; Oudmaijer, R. D.; et al. 2011b, \aap, 534, L10
\bibitem[Lamers \& Cassinelli(1999)]{lam99} Lamers, H. J. G. L. M.  \& Cassinelli, J. P. 1999, Introduction to Stellar Winds (Cambridge University Press)
\bibitem[Le Bertre et al.(1989)]{leb89} Le Bertre, T.; Heydari-Malayeri, M.; Epchtein, N.; et al. 1989, \aap, 225, 417
\bibitem[Lobel et al.(2003)]{lob03} Lobel, A.; Dupree, A. K.; Stefanik R. P.; et al. 2003, \apj, 583, 923
\bibitem[Lobel et al.(2013)]{lob13} Lobel, A.; Groh, J. H.; Martayan, C.; et al. 2013, \aap, 559, 16
\bibitem[Loup et al.(1993)]{lou93} Loup, C.; Forveille, T.; Omont, A. \& Paul, J. F. 1993, \aaps, 99, 291
\bibitem[Mason et al.(1998)]{mas98} Mason B. D.; Gies, D. R.; Hartkopf, W. I.; et al. 1998, \aj, 115, 821
\bibitem[Oudmaijer et al.(1996)]{oud96} Oudmaijer, R. D.; Groenewegen, M. A. T.; Matthews, H. E.; et al. 1996, \mnras, 280, 1062
\bibitem[Oudmaijer(1998)]{oud98} Oudmaijer, R. D., 1998, \aaps, 129, 541
\bibitem[Oudmaijer et al.(2009)]{oud09} Oudmaijer, R. D.; Davies, B.; de Wit, W.-J. \& Patel, M. 2009, in {\it The Biggest, Baddest, Coolest Stars}, ASP Conf. Series 112, pp. 17-32
\bibitem[Raskin et al.(2011)]{ras11} Raskin, G.; van Winckel, H.; Hensberge, H.; et al. 2011, \aap, 526, 69
\bibitem[Reid et al.(2009)]{rei09} Reid, M. J.; Menten, K. M.; Zheng, X. W.; et al. 2009, \apj, 700, 137
\bibitem[Wu et al.(2012)]{wu12} Wu, Y. W.; Xu, Y.; Menten, K. M.; et al. 2012, IAU Symposium, Volume 287, p. 425-426

\end{thebibliography}
\end{document}